\documentclass[aps, prb, reprint, tightenlines]{revtex4-2}

\usepackage{color}
\usepackage{graphicx}
\usepackage{amsmath}
\usepackage{amsfonts}
\usepackage{amssymb}
\usepackage{microtype}
\usepackage{hyperref}
\usepackage{bbm}

\hypersetup{
  colorlinks   = true, 
  urlcolor     = blue, 
  linkcolor    = blue, 
  citecolor   = blue 
}

\usepackage[smallerops]{newtxmath}

\begin{document}

\author{Wouter Buijsman}
\email{buijsman@pks.mpg.de}
\affiliation{Max Planck Institute for the Physics of Complex Systems, 01187 Dresden, Germany}

\author{Pieter W. Claeys}
\affiliation{Max Planck Institute for the Physics of Complex Systems, 01187 Dresden, Germany}

\title{Periodic revivals from supersymmetry in a fermionic kinetically constrained model}

\date{\today}

\begin{abstract}
Supersymmetry provides a natural playground for the construction of dynamically constrained lattice fermion models. We here illustrate how supersymmetry can be used to construct a fermionic equivalent of the PXP model with an adjustable chemical potential. This model is closely related to the $\mathcal{N} = 2$ supersymmetric $M_1$ model, inheriting its integrability. The supersymmetric algebra additionally implies that the dynamics exhibit periodic revivals for specific initial states, including the $\mathbb{Z}_2$-ordered (every second site occupied) product state. These dynamics are reminiscent to those of the PXP model, a paradigmatic effective model in the field of quantum many-body scars. We draw a further parallel by uncovering eigenstates obeying sub-thermal entanglement scaling at energies given by (plus or minus) square roots of integers and relate these to special eigenstates of the $M_1$ model. While we focus on a concrete model, our proposed approach is applicable to more general supersymmetric algebras, where it is expected to lead to non-ergodic dynamics.
\end{abstract}

\maketitle

\section{Introduction}
Starting from the experimental observation of periodic revivals in the dynamics of programmable atom quantum simulators~\cite{Bernien17, Bluvstein21}, it is by now well-established that quantum many-body systems with (emergent) constraints can avoid relaxation up to time scales much longer than naively expected~\cite{Serbyn21, Papic22, Moudgalya22, Chandran23}. Such dynamics can often be related to a small number of highly non-thermal eigenstates known as quantum many-body scars~\cite{Turner18, Turner18-2}, in loose analogy to the single-particle quantum scars first observed by Heller in 1984~\cite{Heller84, Sridhar91, Kaplan99}. These eigenstates occur at finite energy density, thereby strongly violating the eigenstate thermalization hypothesis~\cite{Deutsch91, Srednicki94, Dalessio16}.

Several mechanisms leading to long-lived periodic revivals and the emergence of quantum many-body scars have been identified by now. Among these are the presence of an approximate hidden SU(2) algebra~\cite{Choi18, Schecter19, Bull20, Chattopadhyay20}, a spectrum generating algebra~\cite{Mark20-2, Mark20-3, Moudgalya20, ODea20, Desaules22, Liska23}, the proximity of the model to integrability~\cite{Khemani19, Mark20}, or Bethe-ansatz integrability~\cite{Matsui24, Katsura24}. 
In this work, we focus on a less-studied mechanism that could induce these phenomena in constrained quantum many-body systems: supersymmetry~\cite{Witten82, Gendenshtein85, Cooper95}. As we point out, supersymmetry imposes additional structure on the eigenstates that can not be captured by ordinary conservation laws such as particle number or momentum conservation. For concreteness we consider a model that is additionally integrable and hence non-ergodic, which will allow us to obtain exact results on the eigenstates and the dynamics. However, we emphasize that the presented construction more generally induces non-ergodicity, irrespective of the presence of integrability.

We will consider the supersymmetric algebra underlying the $M_1$ model, a relatively well-studied example of a supersymmetric constrained lattice quantum many-body system~\cite{Fendley03, Fendley03-2}. It describes short-range interacting spinless fermions on a chain (although it can naturally be defined on any graph), subject to the constraint that two neighboring sites can not be occupied simultaneously. The model is integrable by the Bethe ansatz~\cite{Bethe31, Shastry90, Fendley03, Fendley03-2}, and can be realized experimentally~\cite{Yang04, Minar22}. Reference~\cite{Surace20} recently studied weak ergodicity breaking for the $M_1$ model on hypercubic lattice geometries. It was found that a number of highly non-thermal eigenstates, interpreted as quantum many-body scars, can be found thanks to the supersymmetry. In this work, we use the $M_1$ model on a chain as a starting point to construct a fermionic equivalent of the (hard-core bosonic) PXP model. The PXP model~\cite{Fendley04, Lesanovsky12, Turner18, Turner18-2} provides a paradigmatic effective model in studies on quantum many-body scars and weak ergodicity breaking (i.e., the partial breakdown of ergodicity)~\cite{Serbyn21, Papic22, Moudgalya22, Chandran23}. 

The aim of this work is twofold: First, we show how the algebraic structure of supersymmetry can be used to construct kinetically constrained lattice Hamiltonians displaying ergodicity breaking and long-lived periodic revivals. While the procedure holds more generally, we focus on the $M_1$ model for concreteness, for which the resulting Hamiltonian represents a fermionic version of the PXP model with an adjustable chemical potential:
\begin{equation}
H_\text{PXP} = \sum_{i=1}^N P_{i-1} \big( c_i^\dagger + c_i \big)P_{i+1} + \mu F \,,
\label{eq: H-PXP}
\end{equation}
expressed in terms of fermionic creation/annihilation operators $c_i^{\dagger}$ and $c_i$, projectors $P_i = 1-c_i^{\dagger}c_i$, and the total fermion number operator $F$. This model is introduced in detail below. We show how the supersymmetry of the $M_1$ model leads to exact and approximate periodic revivals in the quench dynamics for specific initial states, including the $\mathbb{Z}_2$-ordered (every second site occupied) product state. Periodic revivals for a small set of initial states, typically including the $\mathbb{Z}_2$-ordered product state, are often seen as a hallmark of quantum many-body scarring and have also been observed in the dynamics of the PXP model~\cite{Turner18, Turner18-2} and other models showing weak ergodicity breaking~\cite{Russomanno22}.

Putting the Bethe-integrability of the $M_1$ model into the mix, second, and of independent interest, we identify special eigenstates of the $M_1$ Hamiltonian on a chain. These special eigenstates, occurring at energies given by plus or minus square roots of integers when considering the PXP-like fermion model, support a matrix product state (MPS) description with small bond dimension and a resulting sub-thermal entanglement scaling. Similar special eigenstates at energies given by plus or minus square roots of integers have been observed for the PXP model~\cite{Lin19, Surace21}.

\section{Supersymmetry}
Supersymmetric quantum mechanics provides a versatile tool for the construction of constrained lattice fermion models. An $\mathcal{N} = 2$ supersymmetric quantum model is described by a Hamiltonian of the form
\begin{align}\label{eq:Ham_SUSY}
    H_S = \{Q, Q^\dagger \} = Q Q^{\dagger}+Q^{\dagger}Q \, ,
\end{align}
where the supercharge $Q$ is an operator that squares to zero, $Q^2 = (Q^{\dagger})^2=0$~\cite{Witten82, Gendenshtein85, Cooper95}. The operators $H_S$, $Q$, and $Q^\dagger$ can be seen as the generators of a Lie superalgebra. Specific realizations of this Hamiltonian can be obtained by taking $Q$ and $Q^\dagger$ to be fermionic annihilation and creation operators, respectively, with additional (kinetic) constraints. This Hamiltonian can be directly seen to be positive semi-definite, such that all its eigenvalues are non-negative. Supersymmetry allows for the identification of two classes of eigenstates: All zero-energy eigenstates $| \psi \rangle$ come in singlets obeying $Q | \psi \rangle = Q^\dagger | \psi \rangle =  0$, while eigenstates $| \psi \rangle$ with a strictly positive energy obeying $Q | \psi \rangle = 0$ have degenerate superpartners $Q^\dagger | \psi \rangle$. All eigenstates with strictly positive energy can hence be arranged in such doublets. 

When $Q$ and $Q^{\dagger}$ act as fermionic annihilation and creation operators one can introduce a fermion number operator $F$ satisfying
\begin{align}
    [F,Q] = -Q \,, \qquad [F,Q^{\dagger}] = Q^{\dagger} \,.
\end{align}
Observing that $[F, H_S] = 0$ shows that the fermion number is a conserved quantity. We here propose the study of Hamiltonians of the form
\begin{align}\label{eq:Ham}
    H = (Q + Q^{\dagger}) + \mu F \,,
\end{align}
where $\mu$ is a free parameter having the interpretation of a chemical potential. Such Hamiltonians have recently appeared in Ref.~\cite{Villazon20} in the context of central spin models. As one direct illustration of constraints that supersymmetry imposes on the dynamics of this model, we can consider the evolution of the total fermion number.
While the total number of fermions $F$ is no longer conserved, the dynamics of the fermion number under the dynamics of $H$ are generically non-ergodic. Consider, for example, a state $| \psi \rangle$ with a fixed fermion number, i.e. $F |\psi \rangle = f |\psi \rangle$. This state is a superposition of eigenstates of the supersymmetric Hamiltonian $H_S$ of Eq. ~\eqref{eq:Ham_SUSY} with $f$ fermions. The Hamiltonian $H$ of Eq.~\eqref{eq:Ham} only couples these eigenstates to their superpartners, which have fermion number $f + 1$ or $f-1$. The fermion number of the time-evolved state $| \psi(t) \rangle = e^{-i H t} | \psi \rangle$ is thus highly constrained, since supersymmetry implies that
\begin{align}
   f-1 \leq  \langle \psi(t) |F |\psi(t)\rangle \leq f+1 \, ,
\end{align}
preventing the system from reaching an ergodic steady-state value for which typically $\langle F \rangle$ corresponds to half-filling.

\section{PXP-like fermion model}
As a concrete realization of a supersymmetric lattice Hamiltonian, the starting point of our discussion is the $\mathcal{N} = 2$ supersymmmetric $M_1$ model~\cite{Fendley03, Fendley03-2}.  The $M_1$ model describes spinless fermions hopping on a graph, subject to a nearest-neighbor blockade that forbids two neighboring sites to be occupied simultaneously:
\begin{equation}
Q = \sum_{i} P_{\langle i \rangle} c_i \,, \qquad P_{\langle i \rangle} = \prod_{j \in \langle i \rangle} \big( 1 - c_j^\dagger c_j \big) \, .
\label{eq: Q}
\end{equation}
The sum runs over all sites $i$, and $\langle i \rangle$ denotes the set of sites neighboring site $i$. On a chain, $\langle i \rangle = \{i-1, i+1 \}$ subject to the periodic boundary condition $i + N \equiv i$. The annihilation (creation) operators $c_i$ ($c_i^\dagger$) obey the standard fermionic anticommutation relations $\{ c_i, c_j \} = \{ c_i^\dagger,  c_j^\dagger \} = 0$ and $\{ c_i, c_j^\dagger \} = \delta_{ij}$. The operators $Q$ and $Q^\dagger$ describe the annihilation (creation) of fermions while taking into account the nearest-neighbor blockade constraint, respectively. Because of destructive interference related to the fermionic anticommutation relations, all resulting interactions act locally (nearest-neighbor and next-nearest neighbor). Although the $M_1$ model can naturally be defined on any graph, in what follows we restrict the focus to a chain. The resulting supersymmetric Hamiltonian $H_S$ as given in Eq.~\eqref{eq:Ham_SUSY} is known as the $M_1$ Hamiltonian on the chain and reads
\begin{equation}
H_{M_1} =  \sum_{i=1}^N P_{i-1} \big( c_i^\dagger c_{i+1} + c_{i+1}^\dagger c_i \big) P_{i+2} + \sum_{i=1}^N P_{i-1} P_{i+1}\,.
\label{eq: M1}
\end{equation}
The Hamiltonian consists of the sum of a kinetic part describing constrained hopping and a potential part and we again fix periodic boundary conditions. We obtain the corresponding PXP-like fermion model $H_\text{PXP}$ studied in this work by substituting $Q$ as defined in Eq.~\eqref{eq: Q} on the chain geometry in Hamiltonian of Eq.~\eqref{eq:Ham}, giving the Hamiltonian as given in Eq.~\eqref{eq: H-PXP}. This Hamiltonian can naturally be seen as a fermionic equivalent of the PXP model. Indeed, the PXP model is recovered for $\mu = 0$ when replacing the fermionic operators $c_i^\dagger$ and $c_i$ by their hard-core bosonic analogs.

\section{Periodic revivals}
The dynamics of the PXP model show approximate periodic revivals for specific initial states~\cite{Turner18, Turner18-2}. Periodic revivals for a small set of initial states including, in particular, the $\mathbb{Z}_2$-ordered product state, are often seen as a hallmark of quantum many-body scarring~\cite{Serbyn21, Papic22}.  Here, we show analytically that the quench dynamics of the PXP-like fermion model $H_\text{PXP}$ as given in Eq.~\eqref{eq: H-PXP} exhibits exact and approximate periodic revivals for properly chosen initial product states. Note however that the integrability of this model, which we will discuss below, precludes the notion of many-body scarring since all eigenstates are non-ergodic.

\subsection{Exact revivals} 
Among others, exact revivals can be observed in the dynamics for the $\mathbb{Z}_2$-ordered product state. We denote this state as $| \mathbb{Z}_2 \rangle$, which can be obtained from the vacuum state $| 0 \rangle$ as $| \mathbb{Z}_2 \rangle = c_2^\dagger c_4^\dagger \dots c_N^\dagger |0 \rangle$, where $N$ is taken to be even. The important observation is that this state is an eigenstate of the $M_1$ Hamiltonian as it is annihilated by the kinetic part, having an energy $E = N / 2$ from the (diagonal) potential part, where $N$ is the number of sites. Specifically, we have that
\begin{equation}
H_{M_1} | \mathbb{Z}_2 \rangle = (Q Q^{\dagger}+Q^{\dagger}Q )| \mathbb{Z}_2 \rangle = Q^{\dagger}Q | \mathbb{Z}_2 \rangle = \frac{N}{2} | \mathbb{Z}_2 \rangle \, ,
\end{equation}
where we have used that $Q^{\dagger} | \mathbb{Z}_2 \rangle=0$ because of the blockade constraint. If we now consider the action of the PXP-like Hamiltonian $H_\text{PXP}$ of Eq.~\eqref{eq: H-PXP}, we find that this state is no longer an eigenstate, but rather gets coupled to its superpartner $ Q | \mathbb{Z}_2 \rangle$, which in turn only couples to $ | \mathbb{Z}_2 \rangle$. This Hamiltonian acts on these two states as
\begin{align}
H_\text{PXP}  | \mathbb{Z}_2 \rangle &= Q  | \mathbb{Z}_2 \rangle + \frac{\mu N}{2}  | \mathbb{Z}_2 \rangle, \\
H_\text{PXP} Q | \mathbb{Z}_2 \rangle &= Q^{\dagger}Q| \mathbb{Z}_2 \rangle + \frac{\mu (N-2)}{2}Q   | \mathbb{Z}_2 \rangle \nonumber \\
&= \frac{N}{2}| \mathbb{Z}_2 \rangle + \frac{\mu (N-2)}{2}Q   | \mathbb{Z}_2 \rangle \,.
\end{align}
Taking into account the normalization of $Q | \mathbb{Z}_2 \rangle$, which satisfies $\langle  \mathbb{Z}_2 | Q^{\dagger}Q | \mathbb{Z}_2 \rangle = (N/2)\langle  \mathbb{Z}_2  | \mathbb{Z}_2 \rangle = N/2$, the Hamiltonian $H_\text{PXP}$ given in Eq.~\eqref{eq: H-PXP} of the PXP-like fermion model can be restricted to the subspace spanned by the normalized states $\{ | \mathbb{Z}_2 \rangle, Q | \mathbb{Z}_2 \rangle / \sqrt{N/2} \}$, where it takes the form
\begin{align}
H_\text{PXP} = 
\begin{pmatrix}
 \mu N/2 & \sqrt{N/2} \\
 \sqrt{N/2} & \mu(N-2)/2
\end{pmatrix}\,.
\end{align}
The eigenvalues $E_\pm$ in this subspace directly follow as
\begin{align}
E_{\pm} = \frac{\mu (N-1)}{2} \pm \sqrt{\frac{N}{2} + \frac{\mu^2}{4}}\,.
\end{align}
As such, under the dynamics of Hamiltonian $H_\text{PXP}$ of Eq.~\eqref{eq: H-PXP}, the state $ | \mathbb{Z}_2 \rangle$ will exhibit Rabi oscillations with its superpartner at frequency $\sqrt{2 N + \mu^2}$. In the absence of a chemical potential, i.e. $\mu=0$, the dynamics is particularly simple, and the fidelity $\mathcal{F}(t) = | \langle \mathbb{Z}_2 | e^{-i H_\text{PXP} t}| \mathbb{Z}_2 \rangle|^2$ can be directly calculated as
\begin{equation}
\mathcal{F}(t) = \cos^2\left(\sqrt{\frac{N}{2}}\, t\right) \, ,
\end{equation}
showing perfect revivals with period $\pi / \sqrt{N/2}$. 

The structure discussed here more generally holds for eigenstates of the $M_1$ Hamiltonian. For a doublet of eigenstates $| \psi \rangle$ and $Q |\psi\rangle$ where $|\psi\rangle$ has energy $E$ and fermion number $f$, the Hamiltonian $H_\text{PXP}$ as given in Eq.~\eqref{eq: H-PXP} takes the form
\begin{align}\label{eq: PXP_matrix}
H_\text{PXP} = \begin{pmatrix}
\mu f & \sqrt{E} \\
\sqrt{E} & \mu (f-1)
\end{pmatrix}\,.
\end{align}
As such, starting from an initial eigenstate of the $M_1$ Hamiltonian, the dynamics will consist of periodic oscillations within the two-dimensional subspace of the corresponding doublet with Rabi frequency $\omega = \sqrt{4 E + \mu^2 f^2}$. 
Note that singlets remain eigenstates of the PXP-like Hamiltonian, since they are annihilated by both $Q$ and $Q^{\dagger}$.

\subsection{Approximate revivals}
Next to the exact revivals discussed above, approximate revivals can be observed in the dynamics when the system is initialized in a single-fermion product state. The PXP-like fermion model only couples states with a single fermion, $f=1$, to the vacuum state with no fermions and to states with two fermions, i.e. $f=0$ or $f=2$, following our discussion above. First note that in the single-fermion basis, the $M_1$ Hamiltonian drastically simplifies to a tight-binding Hamiltonian, since in this sector all projector terms evaluate to unity. While the states in the two-fermion basis are more complicated, the relevant eigenstates for the dynamics can be directly obtained from the supersymmetry.

The single-fermion eigenstates of the $M_1$ Hamiltonian are plane waves 
\begin{equation}
|\psi_k \rangle = \frac{1}{\sqrt{N}} \sum_{j=1}^N e^{i k} c_j^\dagger |0 \rangle\,,
\end{equation}
with the momentum $k \in (-\pi, \pi]$ given by an integer multiple of $2 \pi / N$, and the corresponding energy $E_k$ given by
\begin{equation}
E_k = N - 2 + 2 \cos (k) \, .
\end{equation}
Because of the supersymmetry, each of these single-fermion eigenstates forms a degenerate doublet with a two-fermion eigenstate $Q^\dagger | \psi_k \rangle / \sqrt{E_k}$, except for the state with $k=0$, which forms a degenerate doublet with the vacuum since $|\psi_{k=0}\rangle =  Q^{\dagger}|0\rangle / \sqrt{N}$. 

\begin{figure}[!t] 
\includegraphics[width = 0.95\columnwidth]{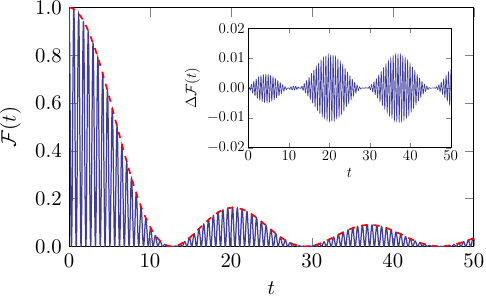}
\caption{The fidelity $\mathcal{F}(t)$ [Eq.~\eqref{eq: fidelity-exact}] for the dynamics of the PXP-like fermion model with $N = 30$ sites initialized in a single-particle product state. The dashed line shows the envelope $J_0^2 (t / \sqrt{N-2})$ appearing in the large-$N$ approximation as given in Eq.~\eqref{eq: fidelity-approx}. The inset shows the difference $\Delta \mathcal{F}(t)$ between the exact fidelity $\mathcal{F}(t)$ of Eq.~\eqref{eq: fidelity-exact} and the large-$N$ approximation as given in Eq.~\eqref{eq: fidelity-approx}. The fidelity exhibits approximate revivals at times scaling as $\sqrt{N}$, and decays as $1/t$ at late times.}
\label{fig: revivals}
\end{figure}

We can now consider the dynamics of an initial single-fermion product state, where we again take $\mu = 0$ for convenience. Such a state is an (in absolute value) equal superposition of all plane-wave single-fermion eigenstates $| \psi_k \rangle$. By the same reasoning as followed above, the fidelity following a quench from a single-fermion product state under the dynamics of the PXP-like fermion model evolves as
\begin{equation}
\mathcal{F}(t) =  \frac{1}{N^2} \left( \sum_k \cos \left[ \sqrt{E_k}\,\, t \right] \right)^2 \,.
\label{eq: fidelity-exact}
\end{equation}
For large system sizes this summation can be approximated by an integral which, along with a Taylor expansion in $1/N$ and the application of basic trigonometric identities, yields the more insightful form
\begin{equation}
\mathcal{F}(t) \approx \cos^2 (\sqrt{N - 2} t) \, J_0^2 \bigg( \frac{t}{\sqrt{N-2}} \bigg)\,,
\label{eq: fidelity-approx}
\end{equation}
where $J_0$ is the Bessel function of the first kind. The fidelity exhibits fast oscillations with period $\pi / \sqrt{N - 2}$ due to the squared cosine term, and slow oscillations with period $\pi \sqrt{N-2}$ due to the oscillations of the Bessel function. The Bessel function additionally induces an envelope that evolves as $1 - t^2 / [2(N-2)]$ at early times, and as $\sqrt{(N-2)/2} / \, t$ at late times. The fidelity hence exhibits a slow power-law decay as $1/t$. Fig.~\ref{fig: revivals} shows a numerical illustration for a system of size $N = 30$. The deviation $\Delta \mathcal{F}(t)$ of the approximation of Eq.~\eqref{eq: fidelity-approx} from the exact result of Eq.~\eqref{eq: fidelity-exact} can be seen to be negligibly small, at least up to moderately late times. We observe from this result that the fidelity of this model exhibits approximate revivals at times scaling with system size as $\sqrt{N}$ for initial single-fermion product states.

\section{Eigenstates and eigenspectrum}
The Hamiltonian $H_\text{PXP}$ of Eq.~\eqref{eq: H-PXP} exhibits additional structure since the $M_1$ Hamiltonian is not just supersymmetric but also integrable. We here first briefly review the exact solution of the $M_1$ model, following the presentation of Refs.~\cite{Fendley03, Fendley03-2}. The Bethe ansatz parametrizes the eigenvalues and eigenstates in terms of parameters $\mu_j$ ($j = 1, 2, \dots, f$). The eigenstate $| \psi \rangle = | \mu_1, \mu_2, \dots, \mu_f \rangle$ corresponding to a set of parameters is expressed in terms of excitations created on top of the vacuum state $| 0 \rangle$ as 
\begin{equation}
| \psi \rangle = \sum_{\{i_1, i_2, \dots, i_f \}} \varphi(i_1,i_2, \dots, i_f) c_{i_1}^\dagger c_{i_2}^\dagger \dots c_{i_f}^\dagger | 0 \rangle
\label{eq: Bethe-psi}
\end{equation} 
with
\begin{equation}
\varphi(i_1,i_2, \dots, i_f) = \sum_P A_P \mu_{P_1}^{i_1} \mu_{P_2}^{i_2} \dots \mu_{P_f}^{i_f} \, .
\label{eq: varphi}
\end{equation} 
The summation in Eq.~\eqref{eq: Bethe-psi} runs over all ordered configurations $( i_1, i_2, \dots, i_f )$ respecting the blockade constraint, thus satisfying $i_{j+1} > i_j + 1$ for all $j = 1, 2, \dots, f-1$. The summation in Eq.~\eqref{eq: varphi} runs over all permutations $( P_1, P_2, \dots, P_f )$ of $( 1, 2, \dots, f )$. The amplitudes $A_P$ and $A_{P'}$ of two permutations $P$ and $P'$ differing only by the exchange of indices $j$ and $k$ are related as
\begin{equation}
\frac{A_P}{A_{P'}} = g(\mu_j, \mu_k) \, ,
\end{equation}
where
\begin{equation}\label{eq:phase_shift}
g(\mu_j, \mu_k) = -\frac{\mu_j (\mu_j \mu_k - \mu_j + 1)}{\mu_k (\mu_j \mu_k - \mu_k + 1)} \,.
\end{equation}
One can view $g$ as the bare scattering matrix describing the phase shift when two excitations are interchanged. Since $g(\mu_j, \mu_j) = -1$, all coefficients are required to be distinct, since otherwise the wave function vanishes. An exception holds for $\mu_j = \exp(\pm i \pi / 3)$, for which both the numerator and denominator vanish. We can define $g(e^{i \pi / 3},e^{i \pi / 3})=g(e^{-i \pi / 3},e^{-i \pi / 3})=1$ without affecting the solvability~\cite{Baxter08}. The phase shift of Eq.~\eqref{eq:phase_shift} more generally simplifies for $\mu_j = \exp(\pm i \pi / 3)$, since
\begin{equation}\label{eq:phase_simple}
g(e^{\pm i \pi / 3}, \mu) = -1/\mu \, , \qquad g(\mu,e^{\pm i \pi / 3}) = -\mu \,.
\end{equation}
With these definitions, the state $| \psi \rangle$ as introduced in Eq.~\eqref{eq: Bethe-psi} is an eigenstate of the $M_1$ Hamiltonian with eigenvalue
\begin{equation}
E = N - 2f + \sum_{j=1}^f \bigg( \mu_j + \frac{1}{\mu_j} \bigg) \,,
\label{eq: Bethe-E}
\end{equation}
provided that the parameters $\mu_j$ form a solution of the Bethe equations
\begin{equation}
\mu_j^N  = (-1)^{f-1} \prod_{k\neq j}^f g(\mu_j, \mu_k) \,.
\label{eq: Bethe}
\end{equation}
The momentum $p$ of an eigenstate is encoded through the relation $e^{ip} = \mu_1 \mu_2 \dots \mu_f$. The Bethe equations directly reflect the underlying supersymmetry of the model. If $\{ \mu_1, \mu_2, \dots, \mu_f \}$ is a solution with all coefficients different from unity, then $\{ \mu_1, \mu_2, \dots, \mu_f, 1 \}$ is a solution with the same energy and momentum, which leads to the  supersymmetry-induced degeneracies between eigenstates with $f$ and $f+1$ fermions.

Additionally, we note that if $\{ \mu_1, \mu_2, \dots, \mu_f \}$ parametrizes a solution with momentum $p$, then $\{ \mu_1^{-1}, \mu_2^{-1}, \dots, \mu_f^{-1} \}$ parametrizes a solution with momentum $-p$. Within the momentum zero and $\pi$ sectors, one can single out solutions for which $\{ \mu_1, \mu_2, \dots, \mu_f \}$ is not equal to $\{ \mu_1^{-1}, \mu_2^{-1}, \dots, \mu_f^{-1} \}$. These solutions are not invariant under spatial inversion, but inversion-symmetric and inversion-antisymmetric eigenstates can be constructed as $|\mu_1, \mu_2, \dots, \mu_f \rangle \pm |\mu_1^{-1}, \mu_2^{-1}, \dots, \mu_f^{-1} \rangle$ by taking the plus or minus sign, respectively. 
The counting of the number of inversion-symmetric and inversion-antisymmetric eigenstates in constrained models has been considered recently in Ref.~ \cite{Buijsman22}.

\subsection{Heisenberg model at root of unity} 
The integrability of this model directly relates to the integrability of the Heisenberg XXZ model since the $M_1$ Hamiltonian can be mapped to the Heisenberg XXZ Hamiltonian under a system-size changing transformation~\cite{Dias00, Fendley03}. This transformation consists of replacing an occupied site together with the left-neighboring empty site by an occupied site, and the remaining empty sites by empty sites. Similar system size changing transformations were also recently discussed in Ref.~\cite{Borsi25}. The resulting Heisenberg Hamiltonian has anisotropy $\Delta = 1/2$ and twisted boundary conditions. This anisotropy corresponds to the Heisenberg model with anistropy at `roots of unity', for which the Bethe ansatz is notoriously subtle~(see, e.g. Refs.~\cite{miao_q_2021,hou_rational_2024} for a detailed discussion). We mainly note that this model exhibits an exponential amount of degeneracies, in addition to the degeneracies originating from supersymmetry, since it supports a representation of the loop algebra $\mathfrak{sl}_2$~\cite{miao_q_2021,hou_rational_2024,deguchi_sl2_2003,deguchi_regular_2007,deguchi_sl2_2001}.

\subsection{Sub-thermal entangled eigenstates} 
The PXP model is known to have exact matrix product eigenstates with system size-independent bond dimensions at energies zero and $\pm \sqrt{2}$~\cite{Lin19}. When imposing the requirement that two occupied sites are separated by at least $n$ empty sites, similar exact matrix product eigenstates at energies zero and $\pm \sqrt{q}$ for a number of integers $q \le n + 1$ have been identified~\cite{Surace21}. Because of the insensitivity of the bond dimension to the system size, these special eigenstates are area-law entangled. Here we show that, in analogy, the PXP-like fermion model hosts eigenstates with energy $\pm \sqrt{q}$ for some integers $q$, which obey sub-volume law entanglement scaling. These states correspond to Bethe states with integer eigenvalues for the $M_1$ model. Various states with integer eigenvalue appear throughout the spectrum of the $M_1$ model, and in Table~\ref{tab: integers} we list these eigenvalues for system size $N=12$ for each fermion number $f$. While such integer eigenvalues have been previously observed in the $M_1$ model~\cite{Razumov01}, we are not aware of any discussion of their origin or the corresponding solutions of the Bethe equations. 

While Bethe states can generally be written as an MPS, the bond dimension typically grows with system size, leading to volume-law entangled eigenstates~\cite{Murg12}. We here identify special classes of Bethe states from the solutions to the Bethe equations as given in Eq.~\eqref{eq: Bethe}, corresponding to integer eigenvalues, and show how these states can be recast as an area-law entangled MPS. Whenever a parameter $\mu_j$ equals $\mu_j = \exp(\pm i \pi / 3)$, this parameter will contribute an integer to the energy $E$ of Eq.~\eqref{eq: Bethe-E} since $\mu_j+1/\mu_j = 1$. 

\begin{table}[t]
\begin{ruledtabular}
\begin{tabular}{l | l l l l l l}
$f$ & $E$ \\ \hline
0 & 12 ($\times$ 1) \\
1 & 8 ($\times$ 1), & 9 ($\times$ 2), & 10 ($\times$ 2), & 11 ($\times$ 2), & 12 ($\times$ 1) \\
2 & 8 ($\times$ 3), & 9 ($\times$ 3), & 10 ($\times$ 2), & 11 ($\times$ 2) \\
3 & 4 ($\times$ 1), & 6 ($\times$ 6), & 8 ($\times$ 2), & 9 ($\times$ 3) \\
4 & 0 ($\times$ 2), & 4 ($\times$ 3), & 5 ($\times$ 5), & 6 ($\times$ 8), & 7 ($\times$ 2), & 9 ($\times$ 2) \\
5 & 4 ($\times$ 2), & 5 ($\times$ 5), & 6 ($\times$ 4), & 7 ($\times$ 2) \\
6 & 6 ($\times$ 2)
\end{tabular}
\end{ruledtabular}
\caption{The integer eigenvalues of the $M_1$ Hamiltonian for system size $N = 12$ given separately for each fermion number $f$. The numbers in brackets denote the number of degeneracies of the eigenvalues. The integer eigenvalues shown here have been identified numerically.}
\label{tab: integers}
\end{table}

We here first focus on special states in which all parameters are identical and satisfy $\mu_j = \exp(\pm i \pi / 3)$. The eigenvalues given in Eq.~\eqref{eq: Bethe-E} can be directly checked to satisfy $E=N-f$. The Bethe equations simplify to a single equation, 
\begin{align}
e^{\pm i \pi N / 3} = (-1)^{f-1} \,.
\end{align}
For system sizes that are an even (odd) integer multiple of 3, this equation is satisfied whenever $f$ is odd (even). The corresponding eigenstates of Eq.~\eqref{eq: Bethe-psi} simplify since
\begin{align}
\varphi(i_1,i_2, \dots i_f)  = \exp\left[\pm i \frac{\pi}{3} (i_1+i_2 + \dots + i_f)\right] \,.
\end{align}
The entanglement of the fermionic wave function follows from $\varphi$ and these states are area-law entangled, as can be made explicit by rewriting them as an MPS with bond dimension $2(f+1)$ (as opposed to volume-law states where the bond dimension scales exponentially with $f$). Specifically, we can write
\begin{align}
\varphi(i_1,i_2, \dots i_f)  = \mathrm{Tr} \left( A_1^{n_1} A_2^{n_2} \dots A_N^{n_N} B\right) \, ,
\end{align}
with $n_i = 1$ if $i \in (i_1,i_2, \dots i_f)$, i.e. the site $n_i$ is occupied, and $n_i = 0$ otherwise. The different matrices are defined as
\begin{align}
    A_j^1 = e^{\pm i \pi j / 3} 
    \begin{pmatrix}
        0 & 1 \\
        0 & 0
    \end{pmatrix} \otimes X \,, \qquad
    A_j^0 = 
    \begin{pmatrix}
        0 & 0 \\
        1 & 1
    \end{pmatrix} \otimes \mathbbm{1} \,, 
\end{align}
and the boundary matrix is given by
\begin{align}
    B = 
    \begin{pmatrix}
        1 & 0 \\
        0 & 1
    \end{pmatrix} \otimes \tilde{B} \,,
\end{align}
where $X$, $\mathbbm{1}$ and $\tilde{B}$ are $(f+1) \times (f+1)$ matrices with $X_{a,b} = \delta_{b,a+1}$ and $\tilde{B}_{a,b} = \delta_{a,f} \delta_{b,0}$. In this construction, the $2 \times 2$ matrices enforce the blockade constraints, where the wave function vanishes if two neighboring sites are occupied, whereas the $(f+1) \times (f+1)$ matrices create the correct number of particles with the appropriate phase.

As apparent from Table~\ref{tab: integers}, the model supports additional eigenstates with integer eigenvalues beyond these states with $E = N-f$ (and their superpartners, which have the same energy but a fermion number differing by one). A large number of these states can be understood as `simple' solutions to the Bethe equations that are dressed by additional parameters $\exp(\pm i \pi /3)$. Consider e.g. $f=1$, for which the dynamics was already argued to reduce to that of a free particle, and the Bethe equation for the single parameter $\mu$ reduces to
\begin{align}
   \mu^N = 1 \quad \rightarrow \quad  \mu = \exp\left(\frac{2\pi i n }{N}\right) \,,
\end{align}
with $n= 0,1 \dots N-1$. The contribution to the energy is given by $2 \cos(2 \pi n / N)$, which takes integer values at various values of $n$. For $N=12$, the integer eigenvalues of Table~\ref{tab: integers} are recovered whenever $n$ is even and for $n=3$ and $9$.

Consider now a solution $(\mu_1, \mu_2, \dots \mu_f)$ to the Bethe equations given in Eq.~\eqref{eq: Bethe} for fixed $N$ and momentum $p$. Then
\begin{align}
    (\mu_1, \mu_2, \dots \mu_f, \underbrace{e^{i \pi/3}, \dots, e^{i \pi/3}}_{n_+} \,,\underbrace{e^{-i \pi/3}, \dots, e^{-i \pi/3}}_{n_-})
\end{align}
provides a solution to the Bethe equations for a lattice of length $N+(n_++n_-)$ with $f+(n_++n_-)$ fermions, provided
\begin{align}
 &\exp\left[i\frac{\pi }{3}(N+n_+)\right] = (-1)^{n_+-1} \exp(-ip) \, , \\
 &\exp\left[-i\frac{\pi }{3}(N+n_-)\right] = (-1)^{n_--1} \exp(ip) \,.
\end{align}
The derivation is straightforward using Eqs.~\eqref{eq:phase_simple}. In this way solutions for $f=1$ with integer eigenvalue can be used to obtain dressed eigenstates with integer eigenvalues at different lattice sizes. These modes with $\mu = e^{\pm i \pi /3}$ act similar to the so-called Fabricius--McCoy strings of the Heisenberg model, which similarly describe excitations that scatter trivially~\cite{fabricius_bethes_2001,fabricius_completing_2001}. 

All nontrivial parameters $\mu_j$ act as excitations that will generally increase entanglement, but for a small value of such nontrivial excitations the eigenstates can still be expected to have sub-thermal entanglement. An eigenstate of the PXP-like fermion model is given by a superposition of an eigenstate of the $M_1$ model and its superpartner. If both eigenstates in this doublet obey sub-thermal entanglement scaling, then so does the eigenstate of the PXP-like fermion model, since this state presents a linear combination of two area-law states. We note that this way of obtaining integer eigenvalues is not exhaustive, only that the entanglement of these states can be directly understood, and refer to Ref.~\cite{hou_rational_2024} for a more detailed discussion of the spectrum.

\section{Conclusions and outlook}
In summary, we studied the dynamics of kinetically constrained models constructed from a supersymmetric algebra. For Hamiltonians of the form of Eq.~\eqref{eq:Ham} the dynamics was argued to be generally non-ergodic, and periodic revivals can be anticipated based on the underlying doublet structure induced by supersymmetry. We illustrated this construction by introducing a fermionic equivalent of the PXP model with an adjustable chemical potential, closely related to the supersymmetric $M_1$ model on a chain. For this model, we established that the quench dynamics for properly chosen initial states display (exact or approximate) periodic revivals, in analogy to what can be observed for the PXP model. The necessary time scales for these revivals can be analytically obtained, and we find that these scale as the square root of system size. Utilizing the Bethe-integrability of the $M_1$ model, we next identified special many-body eigenstates, with eigenvalues at (plus or minus) square roots of integers, that allowing for a matrix product state description with small bond dimension and hence exhibiting sub-thermal entanglement scaling.

Our work provides a number of suggestions for further investigations. It is natural to ask if interpolations between the PXP and PXP-like fermion models provide new insights on the origin of quantum many-body scars in the former. Such interpolations can be obtained, for example, by replacing the fermions in the PXP-like fermion model by proper anyons. The $M_1$ model can also be generalized and deformed in various ways without breaking the supersymmetry and/or the integrability. For example, when replacing the constraint forbidding two neighboring sites to be occupied simultaneously by a constraint forbidding sites up to a distance $k$ apart from an occupied site, an $M_k$ model results~\cite{Fendley03-2}. The $M_2$ model is, like the $M_1$ model, known to be supersymmetric and integrable, while $M_k$ models are supersymmetric for all $k$~\cite{Hagendorf14}. The $M_2$ model has recently gained attention since it additionally corresponds to ``free fermions in disguise"~\cite{Fendley24}. An interesting open question is how our results translate to such generalizations and deformations of the $M_1$ model, and to supersymmetric fermion lattice models in general. Shifting the focus to integrability, this work opens up ways of using the toolbox of integrability to study dynamics in the class of introduced models. Finally, while the eigenstates of this PXP-like fermion model can be exactly expressed as (linear combinations of two) Bethe-ansatz states, it is unclear if and how the conserved charges underlying integrability can be constructed for arbitrary values of the chemical potential.

\acknowledgments
We thank Devendra Singh Bhakuni, Jean-Yves Desaules, and Maksym Serbyn for useful discussions. We thank Zlatko Papi\'c for pointing out the perfect revivals in the discussed model, and Yuan Miao for helpful comments about the eigenspectrum of the Heisenberg Hamiltonian at roots of unity.

\bibliography{PXPLibrary}

\end{document}